\documentclass[aps,prl,reprint,groupedaddress]{revtex4-2}
\usepackage{amsmath}
\usepackage{physics}
\usepackage{slashed}
\usepackage{graphicx}
\usepackage{hyperref}
\hypersetup{hidelinks}
\begin{document}


\title{Decay-Assisted Leptogenesis during Preheating}


\author{Wenbin Zhao}
\email[]{zhaowenbin@ucas.ac.cn}
\altaffiliation{
}
\affiliation{International Centre for Theoretical Physics Asia-Pacific, University of Chinese Academy of Sciences,
100190 Beĳing, China}
\affiliation{Taiji Laboratory for Gravitational Wave Universe, University of Chinese Academy of Sciences (UCAS), Beijing, China.}


\date{\today}

\begin{abstract}
We revisit the non-thermal leptogenesis mechanism during preheating using the quantum Boltzmann equations. This framework allows us to study the production and decay of heavy right-handed neutrinos in a consistent manner. Due to their fermionic nature, right-handed neutrino production can saturate after a few inflaton oscillations when their occupation numbers approach unity. Right-handed neutrino decays can deplete their occupation numbers, allowing more particles to be produced and generating a larger lepton asymmetry. Compared with the case where right-handed neutrinos decay only after production has finished, we find that the final baryon asymmetry can be enhanced by two orders of magnitude.
\end{abstract}


\maketitle
{\bf \textit{Introduction}}---
Leptogenesis is among the most elegant mechanisms to generate the baryon asymmetry of the universe~\cite{Fukugita:1986hr}. It only requires the existence of heavy right-handed neutrinos with CP-violating decays that generate lepton asymmetry. After that, the Standard Model sphaleron processes can convert the lepton asymmetry into a baryon asymmetry~\cite{Kuzmin:1985mm}. These right-handed neutrinos are well motivated particles beyond the Standard Model as they can naturally explain the small neutrino mass through the seesaw mechanism~\cite{Minkowski:1977sc,Gell-Mann:1979vob,Glashow:1979nm,Yanagida:1980xy}.

Depending on the source of right-handed neutrinos in the early universe, leptogenesis can be classified into two categories. In thermal leptogenesis, right-handed neutrinos are produced from scatterings in the hot thermal bath~\cite{Fukugita:1986hr}. Their abundance will be sensitive to the temperature of the thermal bath, leading to a lower bound on the reheating temperature~\cite{Giudice:2003jh}. Meanwhile, active inverse processes can wash out the existing lepton asymmetry, requiring a restricted light neutrino mass window for its successful implementation~\cite{Buchmuller:2003gz}.

Models of non-thermal leptogenesis instead produce right-handed neutrinos directly after inflation, where the reheating temperature can be lower than the right-handed neutrino mass~\cite{Lazarides:1990huy,Asaka:1999yd}. Inflation is a hypothetical period used to explain our almost isotropic, homogeneous universe and generate initial seeds for structure formation~\cite{Guth:1980zm,Mukhanov:1981xt,Bardeen:1983qw,Mukhanov:1990me}. In the simplest scenario, inflation is realized using the potential energy of a scalar field, known as the inflaton. After inflation, the inflaton field oscillates around its minimum and decays into daughter particles to form the hot thermal bath~\cite{Turner:1983he,Kofman:1997yn}. This oscillating background can also produce fermions non-perturbatively, and drive their occupation numbers toward the Pauli bound in a short time~\cite{Traschen:1990sw,Shtanov:1994ce,Greene:1998nh,Greene:2000ew}. This non-adiabatic particle production, known as preheating, is commonly described using mode functions and Bogoliubov transformations. These results have been successfully applied to right-handed neutrinos and leptogenesis~\cite{Giudice:1999fb}, even gravitationally~\cite{Co:2022bgh,Chowdhury:2026zox}. It was realized that including decays can increase the total amount of right-handed neutrinos produced and hence the final baryon asymmetry~\cite{Giudice:1999fb}. Recently, such an enhancement has been shown for an axion-like inflaton with a derivative coupling by resetting the Bogoliubov coefficients to vacuum values between production stages~\cite{Deng:2026fcx}. 

Incorporating additional processes such as decays requires extending the usual collisionless Bogoliubov methods. We therefore adopt quantum kinetic equations to describe the coupled evolution of particle production and decay. Such a description was used to understand the Schwinger pair production in a time-dependent homogeneous electric field~\cite{Schmidt:1998vi}, and subsequently in an  oscillating cosmological background~\cite{Pervushin:2006vh}. Unstable fermion production was derived later using the coherent quasiparticle approximation ~\cite{Herranen:2008hu,Herranen:2010mh}. These works provide the necessary theoretical foundation to understand preheating from a different perspective, using both particle occupation and coherence.

In this Letter, we revisit the non-thermal leptogenesis mechanism during preheating using quantum Boltzmann equations, giving the first unified calculation on how right-handed neutrino lifetime changes the baryon asymmetry across different decay regimes. We find the correct value can be obtained in separated cases, where right-handed neutrino decay becomes relevant during its production, after production but before reheating, or after reheating, through the same equations.

In what follows, we first review the key elements of our scenario, including the inflaton background during preheating, the CP-violating decay of a right-handed neutrino and how to describe its production and decay in an occupation-coherence system. Several limits will be taken to obtain a qualitative understanding of different aspects of the problem. Numerical results regarding the evolution of the system as well as the final baryon asymmetry will be presented and we conclude in the end.

{\bf \textit{Preheating and decay of right-handed neutrinos}}---
We consider a general inflaton potential which features a quadratic minimum around its origin $V(\phi) = \frac{1}{2}m_\phi^2 \phi^2$, where the inflaton mass is taken to be a free parameter. Such a minimum can arise from a broad class of inflationary potentials, including the Starobinsky model~\cite{Starobinsky:1980te}, the no-scale Starobinsky model~\cite{Ellis:2013xoa} and $\alpha$-attractor models~\cite{Kallosh:2013hoa,Kallosh:2013yoa}. In these scenarios, the inflaton field  undergoes harmonic oscillations after inflation. We further assume the inflaton field couples to a Standard Model fermion $f$ and  heavy right-handed Majorana fermions $N_i$:
\begin{equation}
    \mathcal{L}_{\textrm{h}} = g_f \phi \bar{f}f + \frac{1}{2}g_i \phi \bar{N_i}N_i\,
\end{equation}
These couplings lead to the energy transfer from the inflaton field to other fields during the oscillation phase, processes known as preheating and reheating. The former occurs in the early stage of inflaton oscillation and corresponds to non-adiabatic particle production due to their time-varying masses. The subsequent reheating stage can be described using perturbative inflaton decays.

Assuming these right-handed neutrinos are also responsible for the generation of light neutrino masses through the type-I seesaw mechanism, they can decay during  production through the same Lagrangian:
\begin{equation}
    \mathcal{L} = i \frac{1}{2} \bar{N}_i \slashed{\partial}N_i + \frac{1}{2}M_i \bar{N_i}N_i- \left[ y_{ij}\bar{L}_i \tilde{H} N_j + h.c \right].
\end{equation}
where $L_i$ denotes the lepton doublets and $\tilde{H}$ is the conjugate Higgs doublet defined by $\tilde{H} = i \sigma_2 H^*$. After electroweak symmetry breaking, the Higgs field acquires a vacuum expectation value $v = 246\, \textrm{GeV}$, and light neutrino masses are inversely proportional to $M_N$. The complex Yukawa couplings may induce a CP asymmetry in the right-handed neutrino decays through the interference between tree-level and one-loop diagrams~\cite{Covi:1996wh}:
\begin{equation}
    \epsilon_i = \frac{\Gamma(N_i \to l_\alpha H_\beta)-\Gamma(N_i \to \bar{l}_\alpha \bar{H}_\beta)}{\Gamma(N_i \to l_\alpha H_\beta)+\Gamma(N_i \to \bar{l}_\alpha \bar{H}_\beta)}
\end{equation}
We assume a hierarchical right-handed neutrino mass spectrum $M_{1} \ll M_{2,3}$, such that the lightest right-handed neutrino $N_1$ decays are the dominant source of lepton asymmetry. In this case, there is a model-independent constraint on the CP violation parameter in terms of the right-handed neutrino mass and the largest mass difference $\Delta m = m_{\rm max}-m_{\rm min}$ between light neutrinos:
\begin{equation}\label{eq:DI_bound}
    \abs{\epsilon_1} \leq \epsilon_{DI} \equiv  \frac{3}{8\pi} \frac{M_1}{v^2} \Delta m = 9.86 \times 10^{-4} \left(\frac{M_1}{10^{13} \,\textrm{GeV}}\right)
\end{equation}
known as the Davidson–Ibarra (DI) bound~\cite{Davidson:2002qv}. In the second equality we have used $\Delta m \approx 0.05 \,\textrm{eV}$~\cite{Esteban:2024eli}, which comes from the atmospheric neutrino mass-squared splitting measured through neutrino oscillations and assuming a massless lightest neutrino. Hereafter, we write $M \equiv M_1$ for simplicity.

The lepton asymmetry yield $Y_L \equiv n_{\Delta L}/{s}$ in the late universe is controlled by the number of right-handed neutrinos produced, its decay products as well as the entropy of the Standard Model thermal bath. To isolate the contribution from preheating right-handed neutrinos, we take the right-handed neutrino mass $M$ to be much larger than the inflaton mass $m_\phi$. In this case, the perturbative decay of the inflaton field into right-handed neutrinos during reheating is kinematically forbidden. The opposite mass hierarchy has been extensively studied in the literature~\cite{Lazarides:1990huy,Asaka:1999yd,Hahn-Woernle:2008tsk,Zhang:2023oyo}. During the preheating stage, we use the following equation of motion to describe the background:
\begin{equation}
    \begin{split}
        \ddot{\phi} + (3 H+\Gamma_\phi) \dot{\phi} + m^2_\phi \phi  &=0 \,,\\
    \end{split}
\end{equation}
where a dot denotes the derivatives with respect to physical time $t$. For the numerical calculations, we use the same initial conditions as in ~\cite{Peloso:2000hy}. When the inflaton field dominates the energy density of the universe, we can use: 
\begin{equation}\label{eq:inflaton}
    \phi(t)  \approx \sqrt\frac{8}{3}\frac{ M_{\rm pl}\cos(m_\phi t)}{m_\phi t}
\end{equation}
to describe the inflaton solution during inflaton oscillations. $M_{ \rm pl} \approx 2.4 \times 10^{18}\, \textrm{GeV}$ is the reduced Planck mass. Production and decay of right-handed neutrinos in this homogeneous and isotropic background are described by a set of quantum Boltzmann equations~\cite{Pervushin:2006vh,Herranen:2008hu,Herranen:2010mh}. For our real mass system, we adopt the relaxation-time approximation as in~\cite{Shaposhnikov:2008pf, Canetti:2012kh}:
\begin{equation}
\begin{split}\label{eq:Quantum_Boltzmann}
    \dot{f}_k &= W_k \operatorname{Re}(C_k) - \Gamma_k f_k\,, \\
    \dot{C}_k &= - (\Gamma_k + 2i\omega_k)C_k + \frac{W_k}{2}(1-2f_k)\,,\\
    \omega_k &\equiv \sqrt{\left(\frac{k}{a}\right)^2+M_t^2}\,, \quad W_k \equiv \frac{k(\dot{M_t}+H M_t)}{a \omega_k^2}\,.
\end{split}
\end{equation}
where $f_k$ is the occupation of the right-handed neutrino for a given comoving momentum $k$, bounded between 0 and 1. $C_k$ is the canonically normalized, complex coherence function between particle and anti-particle states.  $M_t = M +g \phi$ is the time dependent mass term that sources the production of right-handed neutrinos through $W_k$ and $\Gamma_k = (y^\dagger y)_{11} M_t^2/(8\pi \omega_k)$ is the momentum-dependent decay width. We use $\Gamma_N = (y^\dagger y)_{11} M/(8\pi)$  to denote its vacuum rest-frame value. In the absence of direct coupling $g$, there is still a non-vanishing $W_k$ term, which corresponds to gravitational production of right-handed neutrinos. Our initial conditions are zero occupation and coherence $f_k=C_k=0$. We have omitted inverse processes in the equations.

Let's first consider the case where the decay term can be neglected, which recovers the Bogoliubov limit~\cite{Herranen:2008hu}. According to Eq.\eqref{eq:Quantum_Boltzmann}, a time-varying mass term will source the coherence, which will generate occupations. It is well known that during the preheating stage the production of right-handed neutrinos can be rapid and they could reach the Pauli bound after a few inflaton oscillations~\cite{Greene:1998nh,Greene:2000ew}. The effective creation of the right-handed neutrino occurs when its mass $M_t$ vanishes and the final, phase averaged occupation of the right-handed neutrino reads~\cite{Peloso:2000hy}:
\begin{equation}\label{eq:no_decay}
\begin{split}
    f_k &\approx \frac{1}{2} - \frac{1}{2} \prod_{i=1}^{n_{\textrm{max}}} (1-2 e^{-\pi p_i^2})\,,\\
    p_i &= k/\sqrt{g a^2 \abs{\dot{\phi}}}\,, \quad n_{\textrm{max}} \approx 0.5\times \frac{g M_{\rm pl}}{M}\,.
\end{split}
\end{equation}
where $i$ labels the successive zero crossings of $M_t$. $\dot{\phi}$ measures the inflaton velocity at each zero crossing. The product form follows from the phase-averaged recurrence relation, including Pauli blocking. $n_{\textrm{max}}$ is the total number of zero crossings up to the end of production. As this expression indicates, increasing the coupling $g$ and reducing the right-handed neutrino mass $M$ will prolong the production. Note that $n_{\textrm{max}}$ is independent of the initial amplitude $\phi_0$ as long as $g\phi_0 \gg M$, since the exact value of $\phi_0$ will only affect the first few oscillations. It is also independent of the inflaton mass $m_\phi$, as changing the inflaton mass will simultaneously modify the oscillation frequency and damping of the inflaton amplitude in Eq.~\eqref{eq:inflaton}, which compensate for each other when counting the number of zero-crossings.

Allowing right-handed neutrinos to decay will mitigate the impact of Pauli blocking, allowing more right-handed neutrinos to be produced. When the decay width is large enough such that $\Gamma_k/m_\phi \gg 1$, the occupation and coherence can be completely removed between zero-crossings. In this case, the total, accumulated occupation $N_k$ is the sum of individual productions:
\begin{equation}\label{eq:full_decay}
    N_k \approx \sum_{i=1}^{n_{\textrm{max}}}  e^{-\pi p_i^2}\,.
\end{equation}
It could easily exceed unity, leading to a large enhancement of lepton asymmetry through its subsequent decay. If, at each zero-crossing, right-handed neutrinos are effectively produced, this enhancement factor would equal $n_{\textrm{max}}$. Considering the damping of the inflaton field and limiting the coupling $g$ to leave the inflaton solution intact will weaken this enhancement, but as we will show in the numerical example, this enhancement could still easily reach $\mathcal{O}(100)$. Reducing the decay width means the occupation numbers are only partially depleted between successive production events, leaving a smaller enhancement too. In the opposite limit where $\Gamma_k/m_\phi \ll 1$, the decay process barely affects production. It becomes relevant only when production has ended, when the source term $W_k$ approaches zero. Both occupation and coherence then undergo exponential decay.

Before moving forward, we note that from Eq.~\eqref{eq:Quantum_Boltzmann}, the following inequality holds:
\begin{equation}\label{eq:Pauli_blocking}
\begin{split}
    (1-2f_k)^2+4 \abs{C_k}^2 \leq 1 \,.
\end{split}
\end{equation}
where the equality holds if $\Gamma_k=0$. The occupation number satisfies $0\leq f_k\leq1$, which means the Pauli bound is strictly satisfied in the system. 

Right-handed neutrino decays generate lepton asymmetry and additional radiation, which can be described using 
\begin{equation}
\begin{split}
    \dot{n}_{\Delta L} + 3H n_{\Delta L} &= -\epsilon_{DI} \langle \Gamma_N n_N \rangle_f \\
    \dot{\rho}_r + 4H \rho_r  &= \Gamma_\phi \rho_\phi + \langle \Gamma_N \rho_N \rangle_f  \,,\\
    H^2 =\left(\frac{\dot{a}}{a} \right)^2&=\frac{1}{3 M_{\rm pl}^2}(\rho_\phi+\rho_r + \langle\rho_N \rangle_f) \,.
\end{split}
\end{equation}
where we have neglected wash-out processes as the temperature of the thermal bath is always much smaller than the right-handed neutrino mass. Preheating of the fermion field $f$ also exists. However, in the presence of a perturbative channel, preheating usually leaves its final energy density unchanged ~\cite{Garcia:2021iag}. Once the right-handed neutrino production is over, its evolution can be described similarly:
\begin{equation}
    \dot{\langle\rho_N \rangle}_f+3H{(\langle\rho_N \rangle_f+\langle p_N \rangle_f) = - \langle \Gamma_N \rho_N \rangle_f}
\end{equation} 
where the subscript $f$ denotes a momentum integral weighted by the occupation number:
\begin{equation}
\begin{split}
    \langle\rho_N \rangle_f = \frac{1}{\pi^2 a^3}\int_{0}^{k_{\textrm{max}}} \dd k k^2 \omega_k f_k \,.\\
    \langle p_N \rangle_f = \frac{1}{3\pi^2 a^5}\int_{0}^{k_{\textrm{max}}} \dd k \frac{k^4}{\omega_k} f_k \,.\\
    \langle \Gamma_N n_N \rangle_f = \frac{1}{\pi^2 a^3}\int_{0}^{k_{\textrm{max}}} \dd k k^2 \Gamma_k f_k \,.\\
    \langle \Gamma_N \rho_N \rangle_f = \frac{1}{\pi^2 a^3}\int_{0}^{k_{\textrm{max}}} \dd k k^2 \omega_k \Gamma_k f_k\,.
\end{split}
\end{equation}
In the above equations, we introduced a momentum cut-off $k_{\textrm{max}}$ to regulate the integral.
This lepton asymmetry is essential to generate the baryon asymmetry of the universe through the electroweak sphaleron processes~\cite{Harvey:1990qw}:
\begin{equation}\label{eq:Baryon_asy}
    Y_B =- \frac{28}{79} Y_L \equiv-\frac{28}{79} \frac{n_{\Delta L}}{s}
\end{equation}
where $s = 2\pi^2 g_*/45 T^3$ is the entropy of the Standard Model thermal bath. $g_*$ is the effective number of relativistic degrees of freedom in the thermal bath, which takes the value 106.75 in the Standard Model. To explain the observed baryon asymmetry of the universe, we require $Y_B = 8.7\times10^{-11}$~\cite{Planck:2018vyg}. The entropy of the Standard Model thermal bath can be analytically calculated in two cases. If right-handed neutrino decay happens well before inflaton decay $\Gamma_N \gg \Gamma_\phi$, most of the entropy comes from inflaton decay. We can use the reheating temperature $T_{rh} = (90/(g_*\pi^2))^{1/4} \sqrt{\Gamma_\phi M_{\rm pl}}$ to calculate the entropy. In contrast, when $\Gamma_N \ll \Gamma_\phi$ and  the right-handed neutrino abundance is sufficiently large, the universe will experience a right-handed neutrino dominated period, during which their decays determine the entropy and lepton asymmetry simultaneously. In this case, we have a second reheating temperature $T_{N} \propto \sqrt{\Gamma_N M_{\rm pl}}$ from right-handed neutrino decay. We can estimate the final baryon yield to be 
\begin{equation}\label{eq:right-handed neutrino_dom}
    Y_B = -\frac{28}{79} \epsilon \frac{n_N}{s} \approx 7.0 \times 10^{-11} \left(\frac{\Gamma_N}{10^{-5} \textrm{GeV}}\right)^{1/2}
\end{equation}
where we used the instantaneous-decay approximation to connect the energy density of the right-handed neutrino $\rho_N = M n_N = \rho_r(T_N)$ with the new reheating temperature and substituted the Davidson–Ibarra (DI) bound in Eq.~\eqref{eq:DI_bound}.

{\bf \textit{Results}}---
\begin{figure}[t!]
    \centering
    \includegraphics[width=1\linewidth]{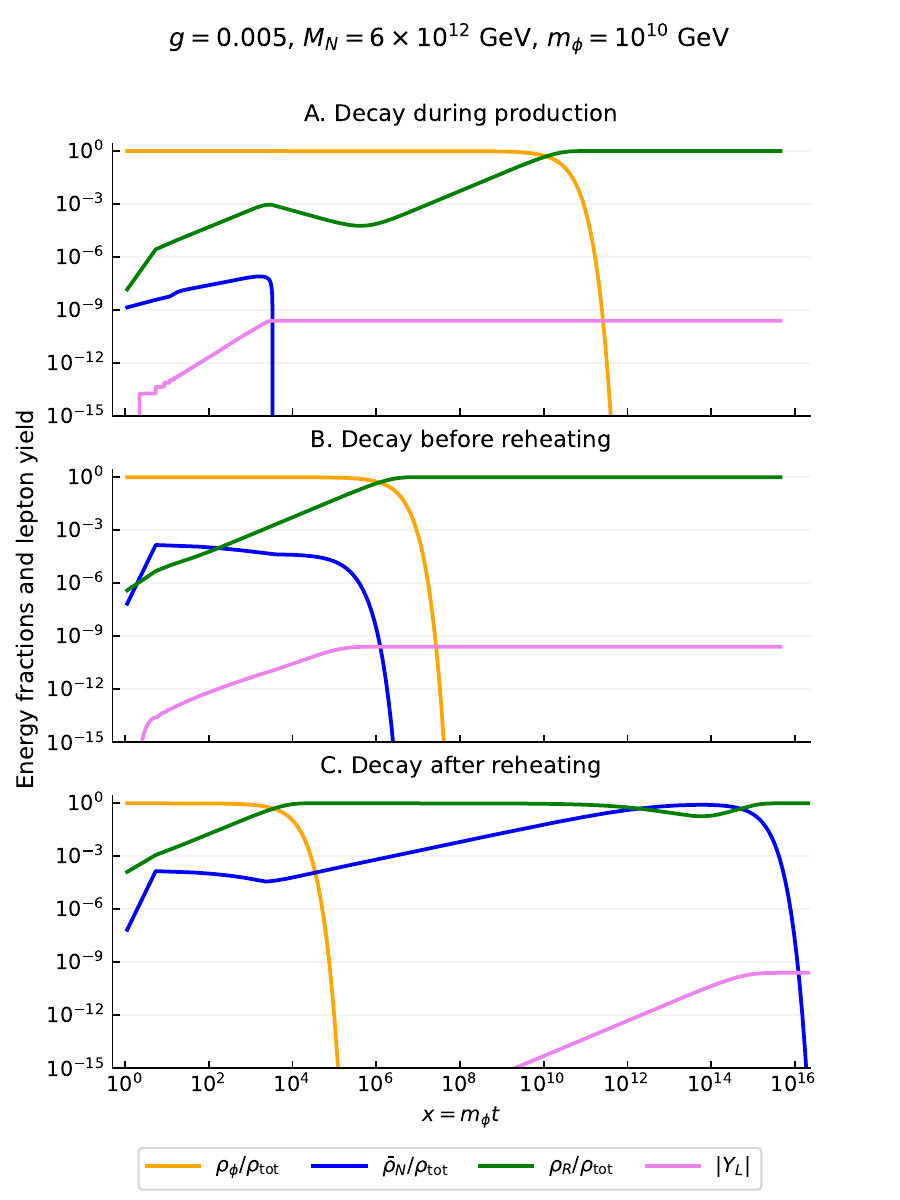}
    \caption{Time evolution of energy densities as well as the absolute value of lepton asymmetry for representative choices of the parameters $\Gamma_N$ and $\Gamma_\phi$. The orange, blue and green lines represent the inflaton, time-averaged right-handed neutrino and radiation energy density, normalized by the total energy density. The pink line shows the lepton asymmetry generated by right-handed neutrino decay, normalized by the final comoving entropy.}
    \label{fig:evolution}
\end{figure}
We have several different parameters in our model which control the time-scales of different processes. The inflaton mass $m_\phi$ is the basic clock which determines the time step for creation of right-handed neutrinos, while the production terminates after approximately $n_{\textrm{max}}$ zero crossings of $M_t$. Additionally, we have $\Gamma_\phi$ and $\Gamma_N$ which control the speed of their damping. Different hierarchies among these parameters lead to different regimes of lepton asymmetry generation. In Fig.~\ref{fig:evolution} we take the following parameters $g=0.005, M = 6\times10^{12} \textrm{GeV}, m_\phi=10^{10} \textrm{GeV}$ and vary $\Gamma_\phi$ and $\Gamma_N$ to illustrate three scenarios. The parameters are chosen such that they give the correct baryon asymmetry after its late time conversion from lepton asymmetry.

Case A represents the scenario where the right-handed neutrino decay width $\Gamma_N \approx 10^{11} \textrm{GeV}$ is larger than both the inflaton mass and decay width $\Gamma_\phi \approx 1 \textrm{GeV}$. Its energy density remains suppressed compared with other cases. Right-handed neutrino decay sources the lepton asymmetry and radiation at the beginning of inflaton oscillations. The radiation energy density fraction first increases due to the rapid decay of right-handed neutrinos, then decreases due to the red-shift and increases again when the inflaton starts to decay. We find the final lepton asymmetry is enhanced by a factor of $100$ compared to the case where $\Gamma_N$ is much smaller, roughly one tenth of the maximal number of zero-crossings $n_{\textrm{max}}\approx1000$.

Case B represents the scenario where the right-handed neutrino decay width $\Gamma_N \approx 10^5 {\textrm{GeV}}$ is smaller than the inflaton mass, but larger than the inflaton decay width $\Gamma_\phi \approx 10^4 \textrm{GeV}$. The right-handed neutrino abundance first increases due to the rapid production from inflaton oscillations, then decreases when the production slows down and high momentum modes redshift. Its decay leads to a long time sourcing of lepton asymmetry and radiation.

Case C represents an extreme scenario where the right-handed neutrino decay width $\Gamma_N \approx 10^{-5} \textrm{GeV}$ is much smaller than the inflaton decay width $\Gamma_\phi \approx 10^6 \textrm{GeV}$. In this case, the right-handed neutrino energy density redshifts more slowly than that of radiation, leading to a late-time increase in the energy fraction. Before decay, it temporarily dominates the energy density of the universe. The final baryon asymmetry hence  depends solely on the right-handed neutrino decay width, as in Eq.~\eqref{eq:right-handed neutrino_dom}.

\begin{figure}
    \centering
    \includegraphics[width=1\linewidth]{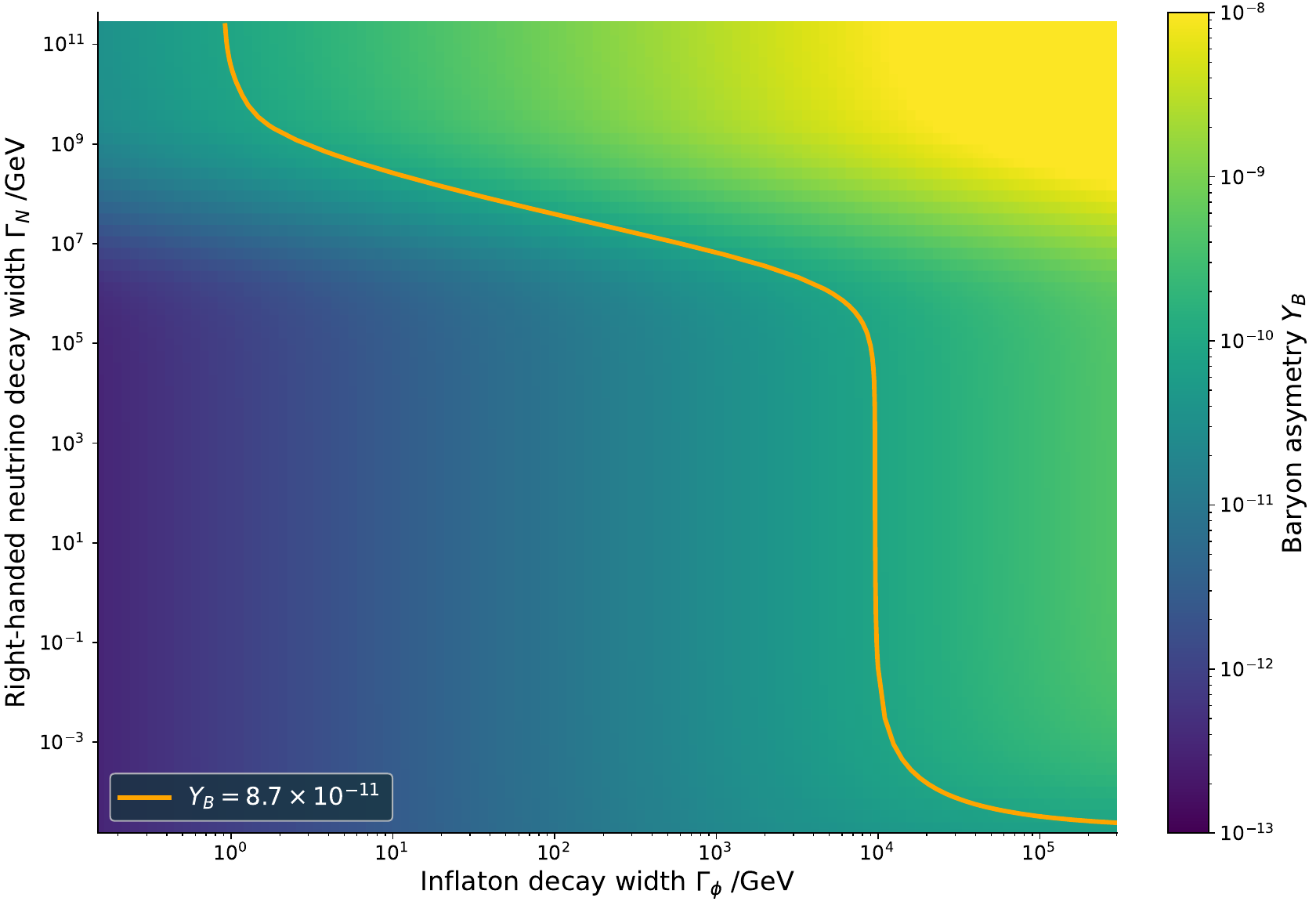}
    \caption{Baryon asymmetry in terms of right-handed neutrino decay width $\Gamma_N$ and inflaton decay width $\Gamma_\phi$. We have fixed the coupling between the inflaton and the right-handed neutrino as well as their masses as in Fig.~\ref{fig:evolution}. The orange line indicates parameter values that reproduce the correct value of baryon asymmetry.}
    \label{fig:contour}
\end{figure}
In Fig.\ref{fig:contour} we show a two dimensional scan over the decay widths $\Gamma_N, \Gamma_\phi$, where the orange curve gives the correct baryon asymmetry in the late time. When $\Gamma_N \gg m_\phi$, as in case A, most of the occupation is removed between two production events. Right-handed neutrino production can be well described using Eq.~\eqref{eq:full_decay},and the final baryon asymmetry is insensitive to the right-handed neutrino decay width.  In the intermediate region $m_\phi / n_{\textrm{max}} < \Gamma_N < m_\phi$, decay is still a relevant process during the production, and the final baryon asymmetry depends on the amount of the occupation being removed. Reducing the right-handed neutrino decay width further $\Gamma_\phi < \Gamma_N < m_\phi / n_{\textrm{max}}$ means the production is intact against the decay process, where the classical result in Eq.~\eqref{eq:no_decay} applies. In this case, the detailed time of right-handed neutrino decay becomes irrelevant in terms of the final baryon asymmetry, as in case B. The last possible choice is $\Gamma_N < \Gamma_\phi$. In this case, the right-handed neutrino decay generates the entropy too, the final baryon asymmetry becomes insensitive to the inflaton decay width $\Gamma_\phi$.

So far our example used a fixed right-handed neutrino mass $M$. Changing it will modify the efficiency of production and hence the possible enhancement of baryon asymmetry. In practice, reducing the right-handed neutrino mass will lead to more production, but also decrease the lepton asymmetry being produced in their decays. Numerically we find these two factors compensate for each other, leading to a relatively stable baryon asymmetry yield. However, when the right-handed neutrino mass is increased, the suppression on production dominates over the asymmetry generation, leading to a decrease of baryon asymmetry.

{\bf \textit{Conclusion}}--- In this paper we have explored a scenario of non-thermal leptogenesis during preheating, taking right-handed neutrino decay into account. Using quantum Boltzmann equations, we showed that different lifetimes of right-handed neutrinos can modify the final baryon asymmetry significantly. With proper parameter choices, we are able to reproduce previous results discussed separately in the literature within a common framework~\cite{Giudice:1999fb}. Assumptions can be relaxed in this work. Using a different inflaton potential could change the preheating results as well as the  background~\cite{Greene:1998nh,Bhusal:2025oqg}, which may lead to a different parameter space for successful leptogenesis. Meanwhile, replacing the Davidson–Ibarra bound with a more accurate estimate taking into account the time-dependent mass of right-handed neutrinos can lead to new insights into our mechanism~\cite{Shuve:2017jgj,ChoeJo:2024wqr}.

We note that our equations can be generalized to a broader range of scenarios, beyond non-thermal leptogenesis. Axions can be produced parametrically and subsequently decay into photons~\cite{Co:2017mop, Carenza:2019vzg, Leedom:2024qgr}. In certain GUTs, supermassive gauge bosons can be produced and subsequently decay to generate baryon asymmetry~\cite{Kolb:1998he}. These phenomenologies can be understood using the same approach by generalizing our fermionic system into bosonic ones.

\begin{acknowledgments}
{\bf \textit{Acknowledgments}}---
This work is supported by the National Natural Science Foundation of China (NSFC) under Grant No.~12547104 and the Fundamental Research Funds for the Central Universities (grant
No.~E5ER6601A2).

\end{acknowledgments}

\bibliography{main}

\end{document}